\documentclass[oldversion]{aa}

\usepackage{epsfig}
\usepackage{graphics}
\usepackage{float}
\usepackage{amsmath}
\usepackage{multirow}
\usepackage{longtable}
\usepackage{rotate}
\usepackage{array}
\usepackage{subfigure}
\def\kms{\ifmmode{\rm km\,s^{-1}}\else\hbox{$\rm km\,s^{-1}$}\fi}

\setlongtables

\begin{document}

\title{The structure of line-driven winds}

\author{L.B.Lucy}
\offprints{L.B.Lucy}

\institute{Astrophysics Group, Blackett Laboratory, Imperial College 
London, Prince Consort Road, London SW7 2AZ, UK}
\date{Received ; Accepted }

\abstract{Following procedures pioneered by Castor, Abbott \& Klein 
(1975, [CAK]), spherically-symmetric supersonic winds for O stars are computed
for matching to plane-parallel moving reversing layers (RL's) from
Paper I (Lucy 2007).
In contrast to a CAK wind, each of these solutions is singularity-free, thus
allowing its mass-loss rate to be fixed by the regularity condition at
the sonic point within the RL. Moreover, 
information propagation in these winds by radiative-acoustic waves is
everywhere outwardly-
directed, justifying the implicit assumption in Paper I that 
transonic flows are unaffected by inwardly-directed wave motions.
}

\authorrunning{Lucy}
\titlerunning{Line-driven winds}
\maketitle

\section{Introduction}

In a recent paper (Lucy 2007, [Paper I]),
models of moving reversing layers (RL's) for O stars were used
to investigate the role of photospheric turbulence in regulating   
the mass flux $J$ that can be accelerated to supersonic velocities. According
to the underlying theory, the integro-
differental equations describing stationary plane-parallel outflow
have $J$ as an eigenvalue, and this is
determined by demanding regularity at the sonic point (Lucy \& Solomon 1970).

The solutions in Paper I were followed to velocities
$v \sim 4a$, where $a$ is the isothermal speed of sound. Left in abeyance,
therefore,  
was the question of matching to solutions
describing the highly supersonic, spherically-symmetric winds where P Cygni
lines are formed. 

Many investigators of line-driven winds follow Castor, Abbott \& Klein
(1975, [CAK]) in using the Sobolev approximation to compute $g_{\ell}$,
the radiation force per gm due to lines. Since this approximation is valid
at the highest velocites reached by the solutions in Paper I,
continuation to $r = \infty$ with a CAK solution would seem to be 
appropriate. But matching a CAK wind to a moving-RL is in general impossible.
Given a star's basic parameters,
namely composition, mass $\cal {M}$, radius $R$ and luminosity $L$, the
regularity condition at the CAK critical point
admits only one solution, with mass-loss rate $\Phi_{CAK}$ as its
eigenvalue. Since the moving-RL already specifies that 
$\Phi = \Phi_{RL} = 4 \pi R^{2} J$, we will in general find that
$\Phi_{CAK} \neq \Phi_{RL}$.

This matching problem is the topic of this paper. At first sight, there are
two approaches. If the moving-RL is the accepted starting point,
then continuation to higher
velocities requires solution of the basic integro-differential system 
since a CAK singularity then no longer exists (Lucy 1975) and so
$\Phi$ is not constrained by a regularity condition at supersonic velocity.   
On the other hand, if the unique CAK solution is the accepted
starting point, then some mechanism must be identified that can force a
structural adjustment of the RL so that $J = \Phi_{CAK}/4 \pi R^{2}$.

\section{Supersonic winds}

This paper's basic premise is that the critical point that determines the
eigenvalue $\Phi$ of a stationary line-driven wind is the sonic point and not
the CAK critical point. In this section, therefore, a model of the
spherically-symmetric supersonic zone is developed for matching to the
plane-parallel sub- and transonic models of Paper I.

\subsection{The model}

Given that the aim is to resolve the paradox of two distict procedures for
predicting $\Phi$, a fairly simple model is adopted: numerical
precision is surely not needed to settle a question of principle. Accordingly,
the supersonic wind is assumed to be isothermal with $T = 0.75 T_{eff}$
and $H$ and $He$ are assumed to be full ionized.

With sphericity now included, the equation of continuity has the integral 
\begin{equation}
  4 \pi r^{2} \rho v = \Phi
\end{equation}
and the equation of motion can then be written as
\begin{equation}
     (v^{2}- a^{2}) \: \frac{1}{v} \frac{d v}{d r} = \frac{2 a^{2}}{r}
                                                + g_{\ell} - g_{*}
\end{equation}
where $g_{*} = (1-\Gamma_{e}) g$.

\subsection{Radiation force}

In Sect. 1, a
plausible argument was given that the calculation of $g_{\ell}$ must be based
on non-Sobolev transfer since otherwise a CAK singularity will be
encountered in the supersonic flow.
Notwithstanding this argument, $g_{\ell}$ is now derived using the Sobolev
approximation. The assumptions are as follows:   

1) As in CAK, $v(r)$ is assumed to be a monotonically increasing
function of $r$, and multi-line scattering is
neglected. Accordingly, each line interacts with dilute but unattenuated
photospheric radiation. 

2) The CAK assumption of radial streaming for photospheric radiation is not
adopted.
Instead, the finite cone occupied by this
radiation is accounted for using modified Eddington approximations
(Lucy 1971). The importance of this departure from CAK was demonstrated by
Pauldrach et al. (1986), but their procedure of introducing this 
effect via a correction factor $CF$ is not followed.    

3) The CAK procedure of approximating the summed contributions of numerous
lines to $g_{\ell}$ by means of an analytic force multiplier function $M(t)$ is
not followed. This approximation has the merit of computational
economy and, more importantly, allows a fairly straightforward analysis of
solution
domains and singular points. But this analysis becomes significantly
more complicated when $CF$ is included, and yet further complicated when the
$M(t)$ parameters vary with radius as in WM-{\em basic} models
(Pauldrach et al. 2001). Accordingly, the force multiplier approach
is dropped and replaced by direct summation.    

The resulting force per gm due to line transitions is
\begin{equation}
  g_{\ell} = \sum_{i} g_{i}
\end{equation}
where $g_{i}$, the contribution from the $i$th line, is given by (Lucy 1971) 
\begin{equation}
  g_{i} = \frac{\kappa_{i}}{c} \: \pi F_{i} \: x^{2} \:
                \frac{1-e^{-\tau_{i,1}}}{\tau_{i,1}}
\end{equation}
Here $\pi F_{i}$ is the flux emitted at the line's
rest frequency $\nu_{i}$ by the photosphere located at $x = R/r = 1$.
Line formation
is accounted for through the factor containing the Sobolev optical depth
given by
\begin{equation}
  \tau_{i,1} = \frac{\kappa_{i} \rho c}{\nu_{i}} \:/
       \left[\mu_{1}^{2} \frac{dv}{dr} + (1-\mu_{1}^{2}) \frac{v}{r}  \right]
\end{equation}
and evaluated at direction cosine $\mu_{1} = \sqrt{1-x^{2}/2}$. The properties
of the transition $l \rightarrow u$ enter through $\kappa_{i}$, given in
standard notation by 
\begin{equation}
  \kappa_{i} \rho = \frac{\pi e^{2}}{m_{e} c} \: n_{l} f_{lu} 
            \left(1- \frac{g_{l} \: n_{u}}{g_{u} \: n_{l}}   \right)
\end{equation}

The above formula for $g_{i}$ 
corresponds to a 1-point quadrature formula for the integration over
$\mu$ using the beam that leaves the photosphere with $\mu_{1} = 1/\surd 2$
and assuming no limb-darkening.

As in Paper I, the data for the $\sim 10^{5}$ lines contributing to $g_{\ell}$
are from Kurucz \& Bell (1995). The fluxes $\pi F_{i}$
are derived from the emergent fluxes of TLUSTY
models (Lanz \& Hubeny 2003) by smoothing with a box filter of width
1000 km s$^{-1}$. Photospheric line-blanketing is therefore included but
wind-blanketing is neglected.

The treatment of ionization and excitation is basically that of
Abbott \& Lucy (1985) as updated in Lucy (1999), with dilution as the dominant
NLTE effect. Chemical composition is as stated in Sect.2.1 of Paper I.

\subsection{The CAK ansatz}

When derived as above, $g_{\ell}(r)$ depends
only on the local properties of the wind at $r$, including the derivative
$v^{\prime} = dv/dr$. But line formation is non-local, occurring over a
finite velocity interval corresponding to a few times the mean
thermal speed of the absorbing ions.
Accordingly, $v^{\prime}$ in Eq. (5) - the Sobolev
derivative - represents an average over this interval. As such, it does not
have the same standing as $v^{\prime}$ in Eq. (2) - the Newtonian derivative.
     
Though doubtless aware of this distinction, CAK nevertheless took the bold step
of assigning equal status to these two derivatives. In consequence,
the right-hand
side of Eq.(2) is a function of $v^{\prime}$, which cannot then  be 
extracted algebraically. 

When making an approximation, even if apparently well justified, we should
be concerned whenever the mathematical nature of the problem changes,
resulting in the addition or removal of crucial aspects of the solutions.  
The Sobolev approximation is certainly plausible for  
high velocity outflows and has been widely and usefully
employed for transfer problems in stellar winds and supernovae.
But when incorporated into
the dynamical theory of winds via the CAK {\em ansatz}, a system of
integro-differental equations is replaced by an ordinary differential
equation (ODE). This results in the following changes to the solutions: 1)
the sonic point is no longer a critical point; and 2) a new (CAK) critical
point appears at supersonic velocities. Since critical points are
certainly crucial, these changes should alert us to the possibility that 
the CAK solution may not be a good approximation to the corresponding solution
of the basic integro-differential system. Despite this reservation, the above
modified CAK theory is used in Sect.3 to investgate supersonic winds.

\subsection{Initial condition}

Given the CAK {\em ansatz}, Eq.(2) reduces to a non-linear first order
ODE with dependent variable $v$. Since the Sobolev
approximation is not well justified in the sub- and transonic flow, an
initial condition $v = v_{1}$ at $x = x_{1}$ is imposed at a point $P_{1}$
where the bulk motion is already highly supersonic.
This point is selected from the appropriate RL of Paper I. 

By thus choosing a supersonic matching point, we also avoid CAK theory's
elimination of the sonic point as a critical point. 
Its correct role appears here implicitly via the initial condition
derived from the RL, where regularity at the
critical point $v = a$ has already been imposed to determine $J$.  

The mass-loss rate $\Phi$ is left as a free parameter
except when set $= \Phi_{RL}$, the rate predicted by the moving-RL.

\section{Solution topology}

In this section, the topology of the solutions of the nonlinear ODE defined
in Sect.2 is investigated. A general analytic discussion such as that of CAK
is not
possible because of the dropping of their simplifying assumptions.  
Instead, the topology is investigated numerically but only along and
in the neighbourhood of the trajectories $v(x; \Phi)$ for specific models.

Following CAK, we write Eq. (2) formally as 
\begin{equation}
  Q \left( v^{\prime}, v, x ; \Phi \right) = 0
\end{equation}
This differential system has singular points on loci defined by 
\begin{equation}
 \frac{\partial Q}{\partial v^{\prime}} = \frac{v^{2} - a^{2}}{v}
                          -\frac{\partial g_{\ell}}{\partial v^{\prime}}  = 0
\end{equation}
Integration through such points is possible only if a regularity
condition is satisfied.

For graphical presentation, it is convenient to make $Q$ and $v^{\prime}$
dimensionless by dividing by $g_{*}(r)$ and $v_{esc}/R$, respectively.
In addition, since $v^{\prime}$ is constrained to be $> 0$
by the assumption of a monotonic velocity law (Sect. 2.2), we define 
$z$ to be the logarithm of the dimensionless $v^{\prime}$.

\subsection{A particular model}

Model S-30 of Pauldrach et al. (2001) is selected for detailed discussion. 
The parameters are $T_{eff} = 30000$K, $log \: g= 3$, and $R/R_{\sun} = 27$,
giving $v_{esc} = \sqrt{2 g_{*} R} = 423$km s$^{-1}$.

According to Paper I, for microturbulent velocity $v_{t} = 10$ km s$^{-1}$,
$J(gm \: s^{-1}) = -5.45$ dex, corresponding to 
$\Phi_{RL} = 2.5 \times 10^{-6} M_{\sun}/yr$. The initial condition
for the integration of Eq.(2) is obtained from the RL
at the point $P_{1}$ where $v = 3a$. Thus $v_{1} = 52.5$ km s$^{-1}$
and this occurs at $x_{1} = 0.950$.

\subsection{Initial derivative}

Because of the dependence of $g_{\ell}$ on $v^{\prime}$, specifying $P_{1}$ 
does not uniquely determine the trajectory $v(x;\Phi)$ for 
$x < x_{1}$ - i.e., the wind's velocity law.
This is illustrated in Fig 1, which
plots the function $Q_{1}(z)$ obtained from Eq.(7) by setting
$(v,x,\Phi) = (v_{1},x_{1},\Phi_{RL})$. We see that the equation $Q_{1} = 0$
has two roots, $z_{R} = 1.81$ and $z_{S} = -0.25$, corresponding to rapid 
and slow initial accelerations, respectively.

\begin{figure}
\vspace{8.2cm}
\includegraphics{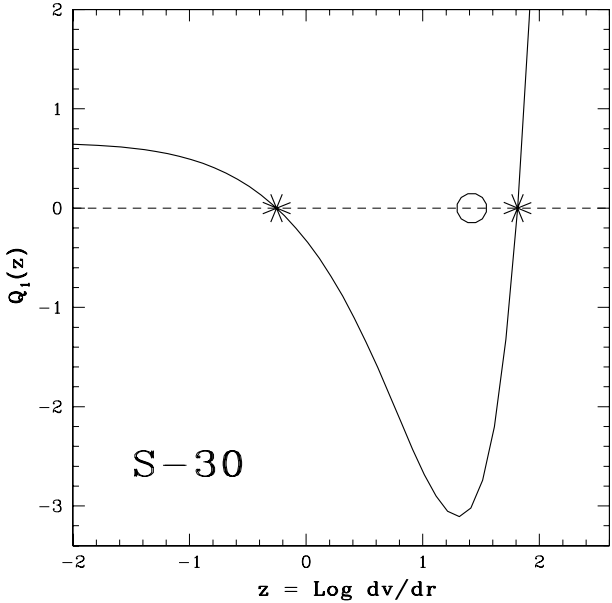}
\caption{Search for velocity gradient at $P_{1}$ for model S-30 
when $\Phi = \Phi_{RL}$. 
The asterisks indicate the two roots of the equation $Q_{1} = 0$. 
The open circle is the prediction
of the moving-RL from Paper I.}
\end{figure}

Fig.2 shows how $z_{R}$ and $z_{S}$ vary with $\Phi$. The roots
coalesce in a double root at $z_{*}=0.78$ when 
$\Phi = \Phi_{*} = 1.05 \times 10^{-5} M_{\sun}/yr$. For $\Phi > \Phi_{*}$,
there is no root and so a matching supersonic solution does not then exist.
On the other hand, for $\Phi < \Phi_{*}$, there are two 
solutions given by the $R$- and $S$-branches. The appropriate continuation is
the one already
selected by the RL solution as it enters the Sobolev
regime. For the S-30 model in Paper I, the velocity derivative at
$v = 3a$ gives $z_{1} = 1.42$ for $\Phi_{RL} = -5.60$ dex, and this point is
plotted in Figs. 1 and 2. Clearly, the point on the $R$-branch is the
appropriate continuation.    

\begin{figure}
\vspace{8.2cm}
\includegraphics{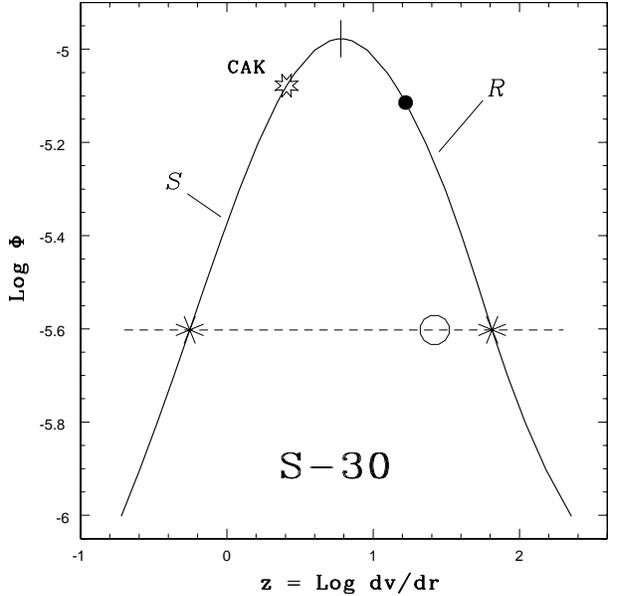}
\caption{Velocity gradients satisfying the equation $Q_{1} = 0$ at $P_{1}$
as functions of $\Phi$. The $R-$ and $S-$ branches are
indicated. The asterisks and the open circle are for 
$\Phi = \Phi_{RL}$, as in Fig. 1. The star on the $S$-branch is
the CAK solution. Points on the $R$-branch below the filled circle give
singularity-free winds extending to $r = \infty$.}
\end{figure}

\subsection{Velocity law}

With the $R$-branch thus selected, Eq. (2) is integrated outwards from
$P_{1}$ in order to determine $v(x;\Phi_{RL})$.
As at $P_{1}$, Eq. (7) continues to have two roots $z_{S}$ and
$z_{R}$ until
$v \simeq 140$ km s$^{-1}$, at which point $z_{S} \rightarrow -\infty$. 
The steeper of the two possible velocity gradients is selected as the natural
continuation of the intial choice of $z_{R}$.  
The integration is continued to $r \ga 200R$, giving
a terminal velocity $v_{\infty} = 1820$ km s$^{-1}$. If the velocity law is
then fitted to the standard form $v = v_{\infty} (1-x)^{\beta}$ at 
$v/v_{\infty} = 0.5$, we obtain $\beta = 0.69$. These results are
unexceptional.

But of exceptional interest is the integration's successful
continuation to $r = \infty$, apparently without encountering a 
singularity.
Because the root-finding uses Newton-Raphson (N-R)
iterations, a convergence failure would occur at a singularity, thus 
terminating the integration at finite $r$.

\subsection{Locus of singular points}

In order to confirm that $v(x;\Phi_{RL})$ does not cross
a locus of singular points, the nearest such points are now found.

When the integration reaches the point
$\vec{\xi_{i}} = (v_{i}, x_{i} , \Phi_{RL})$,
the equation $Q(z_{i}, \vec{\xi_{i}}) = 0$
is solved for $z_{i}$, thus determining $v^{\prime}$ for the next
integration
step. But now $z$ is varied away from $z_{i}$ in order to explore the solution
topology along the line $\vec{\xi} = \vec{\xi_{i}}$ in 
$(v^{\prime},v,x,\Phi)$-space.
Thus we define a sequence of functions 
$Q_{i}(z) = Q (z; \vec{\xi}_{i})$
of the single variable $z$, the first of which $Q_{1}(z)$ is already
plotted in Fig.1.

In Fig.3, this $Q_{i}$-sequence is plotted for a dense set of points $P_{i}$
through the wind, and the stationary points $z^{s}_{i}$ of the $Q_{i}$ are
marked. Since the curve defined by the $z^{s}_{i}$ does not intersect the
line $Q = 0$, these calculations show that nowhere along the supersonic
wind is $\partial Q/ \partial v^{\prime} = 0$, thus confirming the absence of a
singularity and explaining the success of N-R iterations in
extracting $v^{\prime}$.

\begin{figure}
\vspace{8.2cm}
\includegraphics{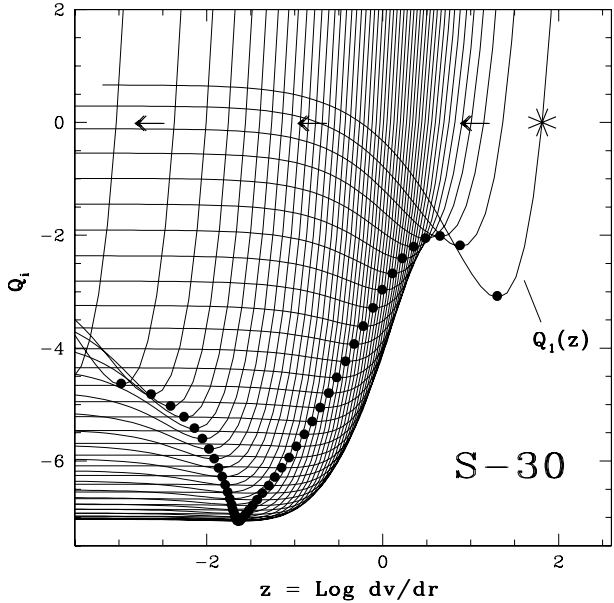}
\caption{Functions $Q_{i}(z)$ for points $(v_{i}, x_{i})$ throughout the
supersonic wind of model S-30. The asterisk is the initial gradient $z_{R}$
and the corresponding $Q_{1}(z)$ is labelled - see also Fig.1. The arrows
indicate the direction of increasing $v$ along the line $Q = 0$ and the filled
circles locate the stationary points $z^{s}_{i}$ of the $Q_{i}(z)$.}
\end{figure}

\subsection{Singularity-free solutions}

Sects. 3.3 and 3.4 show that the moving-RL for S-30
can be matched to a singularity-free
wind extending to $r = \infty$. To explore how special such a continuation
is, a sequence of outward integrations from $P_{1}$ is carried out with
$\Phi$ as parameter. For each integration, the starting derivative
is the $R$-root from Fig. 2, and the integration continues with this root
as in Sect. 3.3. From this sequence, the singularity-free domain
is found to be
$\Phi \leq \Phi_{u} = 7.68 \times 10^{-6} M_{\sun}/yr$. 
This upper limit is indicated on the $R$-branch in Fig.2. The corresponding 
terminal velocity is $(v_{\infty})_{u} = 887$ km s$^{-1}$. 

An integration with $\Phi$ increased to
$7.69 \times 10^{-6} M_{\sun}/yr$ encounters a singularity when
$v = 102.5$ km s$^{-1}$, at which point the N-R iterations fail
to converge. Interestingly, the proximity of a singularity is evident in the
successful integration for $\Phi = \Phi_{u}$ in the form of a sharp decrease
in $dv/dr$ at $v \sim 105$ km s$^{-1}$.
Clearly, $\Phi_{u}$ is the value of $\Phi$ such that the
outward integration just grazes the locus of singular points. The velocity
of this tangent point for S-30 is $v_{u} \simeq 104$ km s$^{-1}$.

Because $\Phi_{u}$ comfortably exceeds $\Phi_{RL}$, the successful continuation
of the RL is {\em not} fortuitous. Such continuations can
therefore be anticipated for RL's covering a wide range
of $v_{t}$.

\subsection{Abbott waves}

With radiative transfer treated using the Sobolev approximation, Abbott (1980)
studied the propagation of density fluctuations in line-driven winds. He
demonstrated that spherically-symmetric radiative-acoustic waves are non-
dispersive and propagate in the local matter frame with velocities
$C_{+}$ and $-C_{-}$, given by 
\begin{equation}
  C_{-} = w + \sqrt{w^{2} + a^{2}} \;\; and
                          \;\; C_{+} = -w + \sqrt{w^{2} + a^{2}}
\end{equation}
where $w = 1/2 \: \partial g_{\ell}/ \partial v^{\prime}$. Transforming
to the star's frame, we see that information can propagate inwards towards
the star if $u < 0$, where 
\begin{equation}
  u = v - C_{-}
\end{equation}

From Eqs. (8)-(10), we readily recover Abbott's major result that
$u = 0$ when $\partial Q/ \partial v^{\prime} = 0$. He thereby demonstated that
the CAK critical point is the furthest point downstream that can communicate
with and therefore influence every other point in the wind. This physical
interpretation of what had hitherto only had a mathematical justification
is undoubtedly a major reason for the almost universal acceptance of CAK
theory as
the correct theoretical foundation for line-driven winds and, in particular, of
the CAK critical point's role in determining $\Phi$.  

Abbott's analysis can be applied to study information flow
in the singularity-free supersonic wind of S-30. Points in
the wind where $u > 0$ cannot communicate with the upstream flow.
From Eqs. (8)-(10), we find that
\begin{equation}
  u > 0 \;\;\ if \;\;\frac{\partial Q}{\partial v{\prime}} > 0 \;\; when
                                       \;\; Q = 0 
\end{equation}
This propagation criterion can be applied to Fig. 3. 
We see that the $R$-root intersections of the
curves $Q_{i}(z)$ with the line $Q = 0$ are all such that $d Q_{i}/dz > 0$.
Thus no point in this solution can 
influence the upstream structure by means of Abbott waves. In particular,
no information propagates back into the RL across the 
boundary at $x_{1}$.  

Fig. 3 suffices for these qualitative remarks. The actual propagation speeds
are as follows: $u=17.4$ km s$^{-1}$ at $P_{1}$,
increases to a maximum of $705.2$ km s$^{-1}$ when
$v = 1644$ km s$^{-1}$, then decreases to $\sim 620$ km s$^{-1}$ in the
terminal flow.

The propagation speeds are also of interest for the limiting singularity-free
solution with $\Phi = \Phi_{u}$ (Sect. 3.5). Initially, $u = 11.4$ km s$^{-1}$
at $P_{1}$, drops to a minimum of $0.8$ km s$^{-1}$
at $v = 104.5$ km s$^{-1}$, then rises monotonically to $217$ km s$^{-1}$
in the terminal flow. This near approach to $u = 0$ reflects this solution's
close proximity to the singular locus when $v \sim 105$ km s$^{-1}$.

The propagation criterion can also be applied to Figs.1 and 2. The $S$-roots
of the equations $Q(z; \Phi) = 0$ are such that $dQ/dz < 0$, so $u < 0$
and information
can propagate back into the RL. The opposite holds for the $R$-roots.
When the roots coalesce, the double root is such that $dQ/dz = 0$, whence
$u = 0$ and $P_{1}$ is a singularity.

\subsection{CAK solution}

As noted above, Abbott demonstrated that $u$ changes sign at $v_{CAK}$, the
velocity of the CAK critical 
point, from being $< 0$ when $v < v_{CAK}$ - i.e., in the upstream sub-critical
flow, to being
$ > 0$ when $v >  v_{CAK}$ - i.e., in the downstream super-critical flow.
Thus, the CAK solution constrained to pass through $P_{1}$ must correspond to
a point on the $S$-branch of Fig.2. Accordingly,
shooting integrations starting at $P_{1}$ with the $S$-root
$z_{S}(\Phi)$ are used
as shown in Fig.4 to
bracket the value of $(\Phi_{CAK})_{1}$. To sufficient accuracy,   
$(\Phi_{CAK})_{1} = 8.35 \times 10^{-6} M_{\sun}/yr$, and this is plotted on
the $S$-branch in Fig.2.

This calculation shows that a CAK-continuation of the moving-RL fails on 
{\em two} counts. Not only is $(\Phi_{CAK})_{1} \neq \Phi_{RL}$ as anticipated
in Sect. 1, but the respective solution branches also differ: the sub-critical
zones of CAK winds are on the $S$-branch, whereas the RL enters the
Sobolev regime on the $R$-branch.

\begin{figure}
\vspace{8.2cm}
\includegraphics{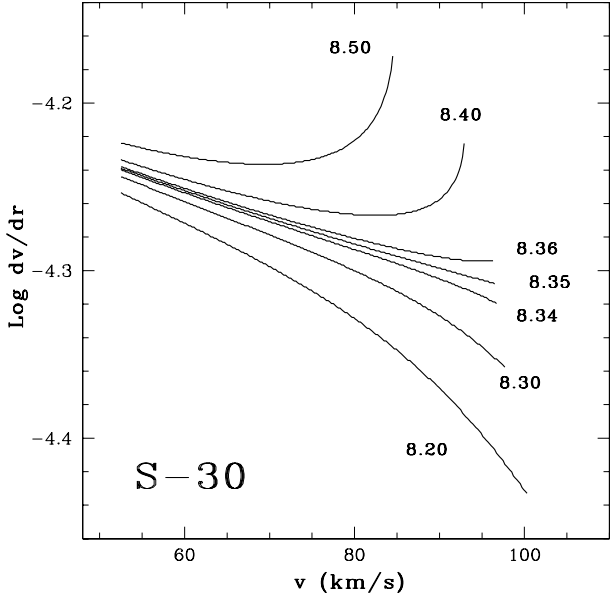}
\caption{Shooting integrations to determine $(\Phi_{CAK})_{1}$, all starting
at $P_{1}$. The curves are labelled with $\Phi$ in units of
$10^{-6} M_{\sun}/yr$.}
\end{figure}

\section{Additional models}

The investigation of S-30 in Sect.3 provides strong support for the approach
in Paper I - i.e, determining
$J$ as an eigenvalue of the equations governing the structure of the
moving-RL, assuming no influence from the highly supersonic
exterior flow.
First, a singularity-free supersonic solution extending to $r = \infty$
exists with $\Phi = \Phi_{RL}$. 
Second, this solution is everywhere super-critical in regard to Abbott waves,
which therefore carry no information back into the RL.

In order to demonstrate that this support is not restricted to S-30, identical
calculations are reported for the two other Pauldrach et al.
models considered in Paper I, namely D-40 and D-50. Numerical details for
all three models are given in Table 1. Velocities are in km s$^{-1}$,
mass fluxes in gm s$^{-1}$ and mass-loss rates in $M_{\sun}/yr$.

\begin{table}

\caption{O-star winds}

\label{table:1}

\centering

\begin{tabular}{c c c c}

\hline\hline

$Quantity$ &  $S-30$  &  $D-40$ &  $D-50$  \\

\hline
\hline

               &       &        &       \\

$T_{eff}$     &  30000     &  40000      &  50000  \\

$log \: g$      & 3.00     &  3.75       &  4.00  \\

$R/R_{\sun}$      & 27     &  10         &  12  \\

$v_{esc}$      & 423       &  743        &  998  \\ 

               &       &        &       \\

$a$         &  17.5        &  20.2       &  22.6  \\

$v_{t}$      &  10         &  10         &  10  \\

$log J$      & -5.45       &  -5.65      &  -4.41  \\

               &       &        &       \\

$x_{1}$     &  0.950       &  0.978      &  0.990  \\

$v_{1}$     & 52.5         &  60.6       &  67.7  \\

$z_{S}$     & -0.25        &  -0.19      &  0.39  \\

$z_{R}$     &  1.81        &  2.64       &  1.82  \\

$z_{1}$ &  1.42            &  1.97       &  1.84  \\

               &       &        &       \\

$\Phi_{RL}$  & $2.5 \times 10^{-6}$ &  $0.22 \times 10^{-6}$  &  $5.4 \times 10^{-6}$       \\

$v_{\infty}$ &  1820       &  5094       &  2578  \\

$\beta$      &  0.69       &  0.61       &  0.63  \\

               &       &        &       \\

$z_{*}$ &  0.78            &  0.90       &  1.06  \\

$\Phi_{*}$  & $1.05 \times 10^{-5}$ &  $2.71 \times 10^{-6}$  &  $1.06 \times 10^{-5}$       \\

               &       &        &       \\

$\Phi_{u}$  & $7.68 \times 10^{-6}$ &  $0.86 \times 10^{-6}$  &  $7.69 \times 10^{-6}$       \\

$(v_{\infty})_{u}$ &  887     &  2442   &  2180  \\

$v_{u}$ &  $\simeq 104$     &  $\simeq 168$    &  $\simeq 189$    \\

               &       &        &       \\

$(\Phi_{CAK})_{1}$  & $8.35 \times 10^{-6}$ &  $1.16 \times 10^{-6}$  &  $8.1 \times 10^{-6}$       \\

$(v_{CAK})_{1}$ &  $\simeq 97$      &  $\simeq 159$      &  $\simeq 156$ \\

               &       &        &       \\

\hline
\hline 

\end{tabular}

\end{table}

\subsection{Comments}

Some noteworthy aspects of Table 1 are as follows:

a) The high $v_{\infty}$ for D-40 suggests that
$\Phi_{RL}$ is too low, but may also be a consequence of the no attenuation
assumption (Sect. 2.2) retained from CAK.  

b) For all three models, $\Phi_{RL} < \Phi_{u}$ so that matching 
singularity-free exterior solutions exist. Moreover, we can anticipate that
this remains true for the {\em reduced} $\Phi_{RL}$'s corresponding to higher 
$v_{t}$'s (see Table 1, Paper I).

c) Comparing the values of $z_{1}$ with the roots $z_{S}$ and $z_{R}$,
we see that in each case the RL selects the $R$-root as the 
appropriate continuation to highly supersonic velocities. The existence of
{\em two} barriers to matching onto a CAK exterior solution is therefore
further supported. 
 
d) In a matched solution, the last point able to communicate with every 
other point in the wind is the sonic point, where $v$
is still only a few percent of $v_{esc}$. This contrasts
dramatically with CAK's O5f model (similar to D-50) where the corresponding
last point $r_{CAK}$ is at $v_{CAK} \simeq 950$ km s$^{-1}$, a velocity well
in excess of $v_{esc}(r_{CAK})$.
Accordingly, no matter moving with $v < v_{esc}(r)$ is out of contact with 
the upstream flow, suggesting perhaps a mechanism that adjusts the
wind's base structure to ensure eventual escape.
But this possibly appealing aspect of the original CAK theory 
does not survive subsequent modifications, especially those of Pauldrach et al.
(1986) - see their Table 1 and Fig. 2a.

e) Because escape is not guaranteed for matter at sonic velocity in the 
RL's, the existence of a matching supersonic solution is essential in
justifying the assumption of stationarity. If there is no such exterior 
solution, we should
expect intermittency due to matter falling back onto the photosphere as well as
departures from spherical symmetry due to a loss of phase
coherence over the stellar surface. 

Moving RL's with no matching supersonic wind will probably be found. However, 
for the three models in Table 1, exterior solutions do exist. The reasons are 
well understood: with only modest changes in $W/n_{e}$, the ratio of dilution
factor to electron density, the ionization balance throughout these winds
is such that matter remains an efficient scatterer of
UV radiation, and Doppler shifts continually irradiate lines with fresh
unattenuated continuum.

\section{Conclusion}

The aim of this paper has been to investigate the problem of matching
the moving-RL's of Paper I to supersonic winds extending to $r = \infty$.
Perhaps surprisingly, matching exterior solutions have been obtained with a
modified CAK procedure using 
the Sobolev approximation. But, in contrast to standard CAK theory,
the entire solution thus constructed has the same critical-point
structure as the basic integro-differential system. Thus, the sonic point
retains its role as a critical point (Paper I), and a CAK critical point no
longer
appears in the supersonic flow (Sect. 3.5). In consequence, there is a 
reasonable
expectation that such solutions will be good approximations to solutions
obtained with non-Sobolev transfer (Sect. 2.3).

Use of a modified CAK procedure 
for the exterior solutions has allowed comparisons
with CAK solution topology and, crucially, with Abbott's discussion of
information propagation by radiative-acoustic waves. With these issues now
clarified, future calculations should re-introduce multi-line scattering.
This can be done by extending the Monte Carlo procedures of Paper I to
$r = \infty$ - see also Abbott \& Lucy (1985) and Vink et al. (2000). Another
attractive possibility is to revive the comoving frame approach
(Mihalas et al.1975,
Weber 1981, Pauldrach et al. 1986). In either case, $\Phi$ is determined
by the constraint of regularity at the sonic point and not at the CAK
critical point, which no longer exists.
Accordingly, scaling relations for $\Phi$ need to be re-investigated
since the basis is undermined for those obtained from the CAK critical point.

\acknowledgement

The referee, S.P.Owocki, is thanked for his comments and suggestions.
This investigation also benefited from points raised by J.Puls in his report on
Paper I.


\begin{thebibliography}{}


\bibitem[]{}
 
 Abbott, D.C. 1980, ApJ, 242, 1183

\bibitem[]{}

 Abbott, D.C., \& Lucy, L.B. 1985, ApJ, 288, 679 

\bibitem[]{}

 Castor, J.I., Abbott, D.C., \& Klein, R.I. 1975, ApJ, 195, 157 (CAK)

\bibitem[]{}

 Kurucz, R.L., \& Bell, B. 1995, Kurucz CD-ROM No.23

\bibitem[]{}

 Lanz, T. \& Hubeny, I. 2003, ApJS, 146, 417

\bibitem[]{}

 Lucy, L.B. \& Solomon, P.M. 1970, ApJ, 159, 879 

\bibitem[]{}

 Lucy, L.B. 1971, ApJ, 163, 95 

\bibitem[]{}

 Lucy, L.B. 1975, Mem.Soc.R.Sci.Li\`{e}ge,8,359

\bibitem[]{}

 Lucy, L.B. 1999, A\&A, 345, 211  

\bibitem[]{}

 Lucy, L.B. 2007, A\&A, 468, 649 (Paper I) 


\bibitem[]{}

 Mihalas, D., Kunasz, P.B., \& Hummer, D.G. 1975, ApJ, 202, 465	


\bibitem[]{}

 Pauldrach,A.W.A., Hoffmann,T.L., \& Lennon,M. 2001, A\&A, 375, 161 

\bibitem[]{}

 Pauldrach, A., Puls, J., \& Kudritzki, R. P. 1986, A\&A, 164, 86	


\bibitem[]{}

 Weber, S.V. 1981, ApJ, 243, 954

\bibitem[]{}

 Vink, J.S., de Koter, A., \& Lamers, H.J.G.L.M. 2000, A\&A, 362, 295


\end{thebibliography}
\end{document}